\documentclass[12pt]{article}
\usepackage{graphicx}
\usepackage{natbib}
\begin{document}

\begin{center}
{\large {\bf
Parameters of galactic disks at optical and NIR wavelengths
}}

\bigskip

A.S.~Gusev$^{1}$\footnote{Email: gusev@sai.msu.ru},
S.A.~Guslyakova$^{2}$, M.S.~Khramtsova$^{3}$

\bigskip

$^1${\it Sternberg Astronomical Institute, Moscow State University,
Moscow 119992, Russia}

$^2${\it Space Research Institute, Russian Academy of Sciences,
Moscow 117997, Russia}

$^3${\it Institute of Astronomy, Russian Academy of Sciences,
Moscow 119017, Russia}

\bigskip

{\small {\it
(Received 1 February 2012)
}}

\bigskip

\end{center}

{\small
We have analyzed the radial scales, central surface brightnesses, and 
colors of 404 disks of various types of galaxies. The central surface 
brightness $\mu_0$ and linear disk scale length $h$ vary smoothly along 
the Hubble sequence of galaxies within a rather narrow interval. The disks 
of relatively early type galaxies display higher central surface 
brightnesses in $K$, higher central surface densities, smaller sizes 
(relative to the diameter of the galaxy), redder integrated and central 
colors. The color gradient normalized to the radius of the galaxy and the 
blue central surface brightness $\mu^0_{0,i}(B)$ of the disk, are both 
independent of the galaxy type. The radial disk scales in different 
photometric bands differ less in early-type than in late-type galaxies. 
The ratio of linear disk scales measured in different photometric bands 
increases with the isophote ellipticity $e$ of the disk (the inclination 
of the galaxy); however, the range of the ratio values for each $e$ value 
exceeds the range of variations of scale lengths ratio over $e$. The disks 
in S0 galaxies have more homogeneous parameters than those in spiral 
galaxies. However, no sharp boundary in the properties of disks in 
lenticular, spiral, and irregular galaxies has been found; all parameters 
vary smoothly along the Hubble sequence. A correlation between the central 
disk surface brightness and the total luminosity of the galaxy is observed. 
We also consider the influence of dust on the photometric parameters of the 
disks. We show that the dust concentrated in dust lines towards the spiral 
arms and bars does not influence to the scale lengths ratio.

\bigskip

}

\section{Introduction}

Knowledge of the photometric parameters of disk components of galaxies is 
essential for studies of the dynamics and evolution of galaxies, dark 
matter, and the distribution of dust in galaxies. In the classical case of 
a thin exponential disk, the disk is described by two parameters in each 
photometric band: the scale length $h$ and the central surface brightness 
$\mu_0$. Various combinations of photometric parameters can be used to 
determine the radial color gradients (which depend on the age and chemical 
composition of the stellar population of the disk and the distribution of 
dust), and to estimate the distribution of the stellar mass and the central 
surface density of the disk.

In numerous studies of the properties of galactic disks
\citep[see the brief review in][]{gusev07}, either galaxies with narrowly 
specified properties (e.g., only Sb galaxies \citep{cunow01}, E--S0 galaxies 
\citep{souza04}, galaxies observed ''face-on'' \citep{jong96}, very 
inclined galaxies \citep{xilouris99}) were considered, or a decomposition 
was carried out using no more than a few (usually, one) photometric bands. 
This substantially reduces opportunities for analyzing the photometric 
parameters of the disks (for example, the absence of IR photometric data 
makes it impossible to determine the influence of dust on the photometric 
properties of a disk). Note also that, in most studies, 1D decomposition 
was based on radial proﬁles of the galaxies, which can in some cases result 
in incorrect values for the disk and bulge parameters \citep{gusev06}.

The photometric properties of disks in the optical and IR are essential for 
a number of important problems, both fundamental and applied. Examples of 
the former include the formation and evolution of disks in S0 galaxies. An 
important applied problem is determining the mass distributions in galaxies 
as a whole, and in their disks in particular. Decomposition of the mass 
distribution into spherical and flat components requires knowledge of the 
distribution of the radiation in the disk and bulge of the galaxy. In 
general, the surface density is proportional to the surface brightness of 
the old stellar population corrected for absorption by dust. In a first 
approximation, it is assumed that the mass distribution corresponds to the 
radiation distribution in the $K$-band. However, IR observations are 
currently much more complicated (and rarer) than $BVRI$ observations. 
Unfortunately, the data from the 2MASS Catalog ($JHK$ photometry) cannot be 
used to study the weak outer regions of galaxies. We have attempted to 
establish a relationship between the radiation distributions (disk scales) 
in optical bands and in the IR.

Here, we use a sample of 404 galaxies with various morphological types and 
wide ranges of disk inclinations and galaxy luminosities to study the 
dependence of the photometric parameters of the disks on the morphological 
type of the galaxy, radial variations of the disk color indices, the 
influence of dust on the photometric parameters of the disks, and the 
dependence of the observed scale for the radial brightness decrease on the 
disk inclination.

\section{The sample of galaxies}

For our study of the photometric parameters of disks in galaxies of various 
types, we used the data from the literature for 392 galaxies 
\citep[see][]{gusev07}, together with our previously obtained CCD photometry 
for 12 galaxies \citep{gusev07,bruevich10}. The integrated parameters of the 
galaxies (including the ellipticity $e$ of the disk isophotes) were taken or 
calculated from data in the LEDA Catalog. We used the derived integrated 
parameters to determine the disk scale lengths $h$ in kpc and the disk 
central surface brightness $\mu^0_{0,i}$ in mag/arcsec$^2$, corrected for 
the inclination of the galaxy. Thus, by using the available single-source 
data for the integrated parameters of the galaxies (taken from the LEDA 
database), we were able to decrease the dissimilarity of the sample objects. 
In addition to the overall sample of 392 galaxies, we considered a separate 
subsample containing the 144 galaxies studied in 
\citet{cunow01,souza04,jong96,xilouris99,mollenhoff04} via 2D 
decompositions. We believe that the disk parameters determined in these 
studies are more reliable than those obtained via 1D decompositions
\citep{gusev06}.

We also used our previous multi-color CCD photometry data of 12 galaxies. 
A description of the data reduction is given in 
\citet{gusev07,gusev06,bruevich10}; the technique used for the 2D 
decompositions of the galactic radiation into bulge and disk components is 
presented in \citet{gusev06,bruevich10}.

\begin{table*}
\begin{center}
\caption{Basic data on the galaxies.}
\label{table:tab1}
\begin{tabular}{rcrrrrrcc}
\hline
 NGC & Bands & Type & $M_B^{0,i}$ & $D$, & $R_{25}$, &
$V_{\rm rot}$, & $e$ & $M_{\rm dust}$, \\
     &       &      &             & Mpc  & kpc       &
km/s           &     & $10^6 M_{\odot}$ \\
\hline
  524 & $U...K$ & -1.2 & -21.63 & 32.4 & 17.0 & 300 & 0.05 & 0.35 \\
  532 & $U...K$ &  2.0 & -19.48 & 31.5 & 16.0 & 191 & 0.74 & 3.3 \\
  783 & $U...K$ &  5.1 & -21.14 & 70.5 & 16.8 &  46 & 0.25 & 26 \\
 1138 & $U...K$ & -2.1 & -19.57 & 32.9 &  8.7 &  25 & 0.05 & --- \\
 1589 & $U...K$ &  1.8 & -21.73 & 49.5 & 23.8 & 323 & 0.63 & 1.3 \\
 2336 & $U...K$ &  4.0 & -22.32 & 32.2 & 30.0 & 256 & 0.42 & 9.7 \\
 4136 & $B...K$ &  5.3 & -18.41 &  7.6 &  4.1 &  93 & 0.18 & 0.17 \\
 5351 & $B...K$ &  3.1 & -21.19 & 48.9 & 19.6 & 202 & 0.53 & 1.3 \\
 5585 & $U...I$ &  6.9 & -18.48 &  5.7 &  3.5 &  79 & 0.38 & 0.12 \\
 7280 & $U...K$ & -1.0 & -19.41 & 25.9 &  8.1 & 131 & 0.36 & 0.056 \\
 7721 & $U...I$ &  4.9 & -21.14 & 26.3 & 11.6 & 142 & 0.75 & --- \\
I1525 & $U...I$ &  3.1 & -21.85 & 69.6 & 19.7 & 186 & 0.31 & --- \\
\hline
\end{tabular}
\end{center}
\end{table*}

Table~\ref{table:tab1} presents the basic data for the 12 added-galaxies: 
bands of the observations, the type, absolute magnitude $M_B^{0,i}$ corrected 
for absorption in the Galaxy and due to the disk inclination, distance to the 
galaxy $D$ in Mpc, radius of the galaxy $R_{25}$ in kpc determined from 
the $25^m$/arcsec$^2$ $B$ isophote, maximum rotational velocity $V_{\rm rot}$ 
in km/s corrected for the inclination, disk isophote ellipticity $e$, and 
mass of dust $M_{\rm dust}$ in solar masses. Most part of parameters were 
taken from LEDA electronic database \citep[see][]{gusev07}.

The disk photometric parameters $h$ and $\mu_0$ for the 12 galaxies in the 
various filters are presented in \citet{gusev07,bruevich10}.

\begin{figure*}
\centerline{\includegraphics[width=10cm]{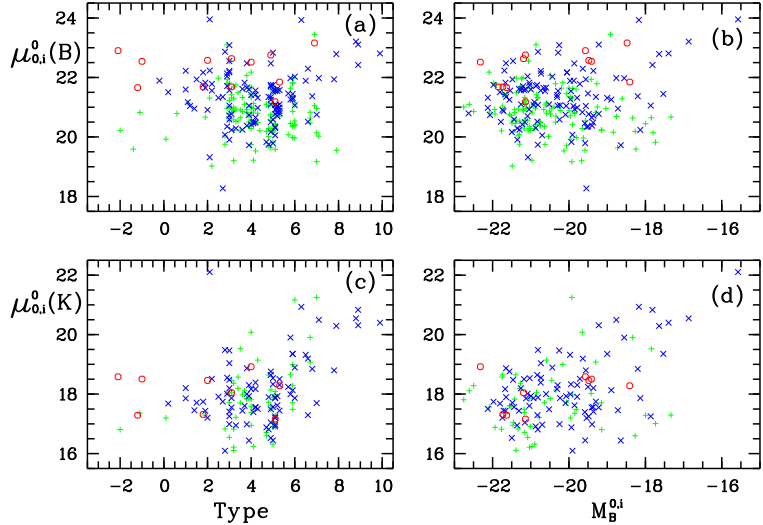}}
\caption{The central disk surface brightnesses $\mu_{0,i}^0$ in $B$ 
(a, b), and $K$ (c, d) as a function of the type (a, c) and absolute 
magnitude $M_B^{0,i}$ (b, d) of the galaxy. Circles mark objects from the 
sample of 12 galaxies, diagonal crosses those from the sub-sample of 144 
galaxies, and the vertical crosses the remaining objects from the sample 
of 392 galaxies.
\label{figure:f1}}
\end{figure*}

\section{Analysis of the results}

\subsection{Central surface brightnesses and color indices of the disks}

In spite of the fact that the galaxies considered have a large range of 
sizes ($R_{25} = 2 −- 40$ kpc) and luminosities ($L_{max}/L_{min}=100$), 
the central surface brightnesses of all the galactic disks $\mu^0_{0,i}$ 
lie in a fairly narrow interval, from $20.2^m$/arcsec$^2$ to 
$22.7^m$/arcsec$^2$ in $B$ and from $16.9^m$/arcsec$^2$ to 
$19.3^m$/arcsec$^2$ in $K$. The central disk surface brightnesses of the 
galaxies considered differ by no more than an order of magnitude. Note, 
there is no dependence between the central surface brightness $\mu^0_{0,i}$ 
corrected for the inclination and the disk inclination itself.

Let us consider the dependence between the central disk surface 
brightnesses in various photometric bands and the luminosity and type of 
the galaxy (Figs.~\ref{figure:f1}a–-\ref{figure:f1}d). In spite of the 
large scatter in the corresponding plots, a correlation between 
$\mu_{0,i}^0$ and $M_B^{0,i}$ is observed, as well as a correlation 
between $\mu_{0,i}^0$ and the galaxy type in long-wavelength bands. 
The luminosities and types of the galaxies in our sample are weakly 
correlated: since there are no bright Sd–-Irr galaxies, it is 
difficult to determine which of the parameters (the luminosity or 
type) is influencing $\mu_{0,i}^0$. Considering the sample of 
144 galaxies whose disk parameters were derived from 2D decompositions, 
we found that the $B$ and $K$ central disk surface brightnesses 
in the early-type galaxies (S0–-Sc) are independent of 
$M_B^{0,i}$ and the morphological type (the correlation coefficient 
$|r| < 0.3$). However, the central disk surface brightnesses in the late 
type galaxies (starting from Sc) increase with the integrated luminosity 
and depend on the morphological type. Considering the galaxies of all 
morphological types, we obtained the dependence 
$\mu_{0,i}(K) \sim (0.16\pm0.06)T$ ($r = 0.45$). $\mu_{0,i}(B)$ is 
independent of the galaxy type (Fig.~\ref{figure:f1}a). Thus, on average, 
the central red brightness of the disk decreases from earlier to later galaxy 
type, while the central blue brightness remains constant for galaxies of all 
morphological types (Figs.~\ref{figure:f1}a,~\ref{figure:f1}c). This 
conclusion seems controversial, and may be a consequence of our selection 
of the objects. We can only suggest that the central color indices of the 
disks depend on the galaxy type.

\begin{figure*}
\centerline{\includegraphics[width=10cm]{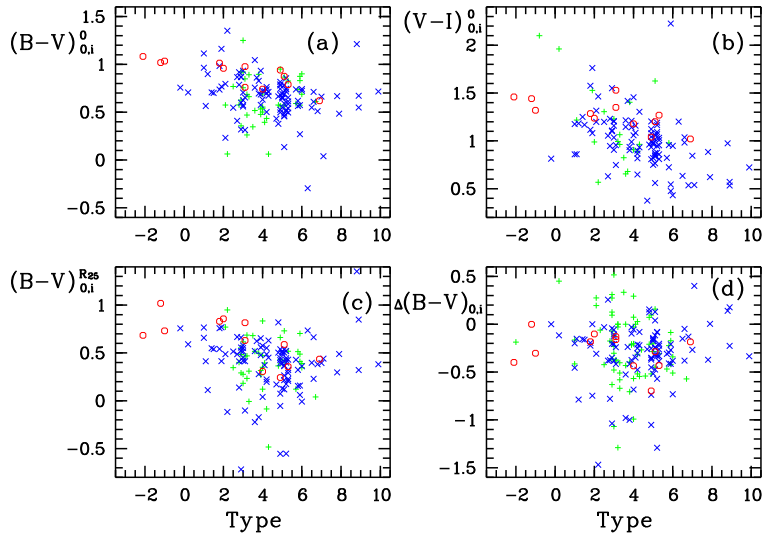}}
\caption{The central disk color indices (a) $(B-V)_{0,i}^0$ and 
(b) $(V-I)_{0,i}^0$, (c) $(B-V)_{0,i}^{R_{25}}$ color indices at the disk 
edges, and (d) radial $\Delta(B-V)_{0,i}$ color gradient as functions of 
the galaxy type. Notation is the same as in Fig.~\ref{figure:f1}.
\label{figure:f2}}
\end{figure*}

Figures~\ref{figure:f2}a,~\ref{figure:f2}b present the dependence between the 
central disk color indices $(B-V)_{0,i}^0$ and $(V-I)_{0,i}^0$ and the galaxy 
type. The centers of the disks are redder in early type than in late type 
galaxies for all three samples and for both shown color indices. Similar 
dependences are derived for the other color indices, with the exception of 
$J-H$ and $H-K$.

The color indices at the disk edges (at a distance $R_{25}$ from the center 
of the galaxy), calculated using the formula 
$(X-Y)_{0,i}^{R_{25}} = (X-Y)_{0, i}^0+1.086R_{25}[1/h(X)-1/h(Y)]$, 
show the same dependence on the galaxy type as the disk centers 
(Fig.~\ref{figure:f2}c). Thus, it appears that the radial gradient of the 
disk color $\Delta(B-V)_{0,i} = (B-V)_{0,i}^{R_{25}}-(B-V)_{0,i}^0$ does not 
depend on the morphological type of the galaxy (Fig.~\ref{figure:f2}d): most 
disks in galaxies of any type become bluer towards their periphery.

No correlation was found between the color indices of the disks and the 
luminosity and inclination of their galaxies.

\begin{figure*}
\centerline{\includegraphics[width=10cm]{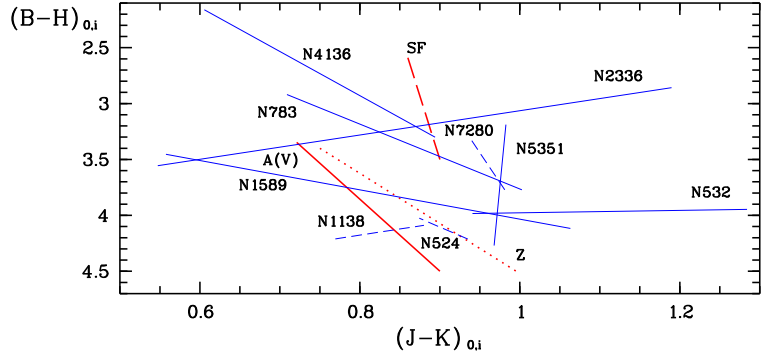}}
\caption{$(B-H)_{0,i} - (J-K)_{0,i}$ two-color diagram for the disks. The 
bold solid curve indicates the displacement of the points due to absorption 
by dust, by $A_V = 1.0^m$ up and to the left. The dotted curve shows the 
metallicity gradient for a stellar system with an age of 10~Gyrs 
\citep[according to][]{bothun90}. The bold dashed curve represents the 
displacement of the points in the case of a burst of star formation 
\citep{bothun90}. Systems with higher metallicity display higher $J-K$ 
color indices. The thin solid curves indicate the radial color variations 
in the disks of spiral galaxies, and the dashed curves the radial color 
variations in the disks of S0 galaxies (for objects from the sample of 9 
galaxies). The NGC numbers of the galaxies are marked in the graph.
\label{figure:f3}}
\end{figure*}

Two-color diagrams may help us qualitatively estimate the composition of 
the stellar population of the disks, as well as study the star formation 
history and the impact of dust in the disks. The interpretation of the 
results derived from optic two-color diagrams can be ambiguous, since 
variations of the metallicity/stellar ages and selective absorption by 
dust both shift the points in the same direction --– along the normal 
color sequence for the integrated colors of galaxies. Two-color diagrams 
with IR color indices can be used to discriminate between the effects of 
dust and age or chemical composition variations.

Figure~\ref{figure:f3} presents the $(B-H)_{0,i} - (J-K)_{0,i}$ diagram for 
the galaxies. According to the models of \citet{bothun90}, variations of the 
age of the stellar population primarily affect $B-H$, while variations of the 
metallicity and absorption by dust primarily affect $J-K$. Unfortunately, 
the photometric parameters of the disks have never been studied 
simultaneously in the $BJHK$ bands, so that we can apply this test only to 
the 9 galaxies observed by us. Note that the straight-line sections 
characterizing the radial variations of the disk color indices are 
appreciably shorter for the S0–-Sa galaxies than for the spiral galaxies 
(Fig.~\ref{figure:f3}). This provides evidence for small radial variations in 
the stellar population and a weak influence of dust in lenticular galaxies. 
All the spiral galaxies (with the exception of NGC~5351) display strong radial 
gradients of their metallicities; it is also possible that absorption by 
dust increases with distance from the center. In late type (Sbc--Scd) spiral 
galaxies, the average age of the disk stellar population also typically 
decreases, and the lower the average age of the disk, the larger the 
gradient for the decrease in age from the center to the edge of the galaxy.

\subsection{Absolute and relative size of the disks}

Both the absolute and relative sizes of the galactic disks lie in even 
narrower intervals than those for the central surface brightness. For the 
vast majority of galaxies, the absolute length scales for the disks are 
2–-7 kpc, while their relative scales are $h/R_{25} = 0.20-0.40$. Note 
that the scatter in $h$ and $h/R_{25}$ substantially exceeds the difference 
between the absolute and relative disk scales measured in different filters 
(except for the $U$ and $B$ bands). On average, the measured $h$ and 
$h/R_{25}$ values decrease from $U$ to $K$.

\begin{figure*}
\centerline{\includegraphics[width=10cm]{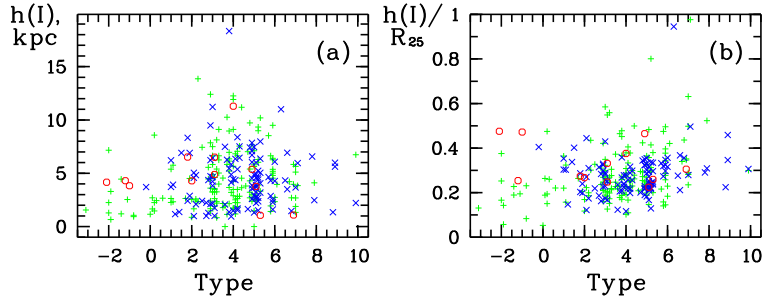}}
\caption{Dependence of the (a) absolute $h$ and (b) relative $h(I)/R_{25}$ 
sizes of the disk on the galaxy type. The notation is the same as in 
Fig.~\ref{figure:f1}.
\label{figure:f4}}
\end{figure*}

The linear disk scale is virtually independent of the morphological type 
of the galaxy (Fig.~\ref{figure:f4}a). Any variation of $h(I)$ with galaxy 
type is substantially smaller than the range of $h(I)$ for each given 
morphological type. Note that the $h(I)$ range in S0 and Sd-–Irr galaxies is 
half that in spiral galaxies: 1–-6 kpc vs. 1–-12 kpc (Fig.~\ref{figure:f4}a). 
With only one exception, the linear scales for the disk brightness decreases 
in S0 galaxies do not exceed 5 kpc; note that our sample contains S0 galaxies 
with both moderate and high (for example, NGC~524) luminosities. 

The plot of $h(I)/R_{25}$ as a function of the galaxy type 
(Fig.~\ref{figure:f4}b) appears more informative than the plot in 
Fig.~\ref{figure:f4}a. Clearly, the later the morphological type of the 
galaxy, the larger, on average, the relative size of its disk. For S0 
galaxies, $h(I)/R_{25} \approx 0.15-0.25$, while for Sd–-Irr galaxies, 
$h(I)/R_{25} \approx 0.30-0.40$. Similar dependences are obtained for the 
absolute and relative disk sizes measured in the other photometric bands.

\subsection{Ratio of linear disk scales in various photometric bands}

\begin{figure*}
\centerline{\includegraphics[width=10cm]{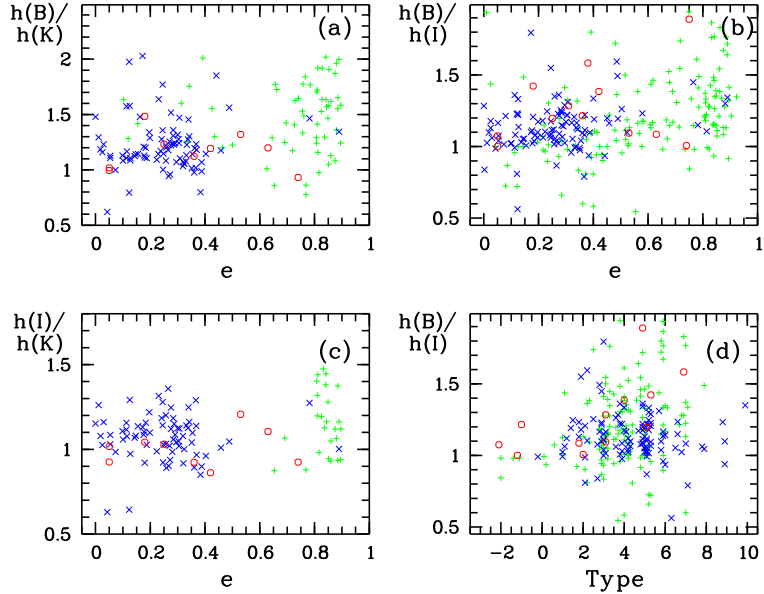}}
\caption{Dependence of the ratios of the radial disk scales (a) 
$h(B)/h(K)$, (b, d) $h(B)/h(I)$, and (c) $h(I)/h(K)$ on the (a–-c) isophote 
ellipticity $e$ and (d) galaxy type. Notation is the same as in 
Fig.~\ref{figure:f1}.
\label{figure:f5}}
\end{figure*}

The parameter $h(X)/h(Y)$, where $X$ and $Y$ are different photometric 
bands, is most sensitive to the presence of dust, and the dependence of 
$h(X)/h(Y)$ on the isophote ellipticity $e$ forms the basic observational 
data for determining the parameters of dust disks in galaxies 
\citep{cunow01}. It was shown in \citet{gusev07} that the average disk 
scale ratios obtained by different authors differ strongly, as do the 
dependences of $h(X)/h(Y)$ on $e$. Using the total sample, we obtained 
relatively small average $h(X)/h(Y)$ values. The values for $h(B)/h(I)$ 
and $h(B)/h(K)$ agree with the estimates of \citet{grijs98} for the case 
when there are appreciable effects only due to the radial age and 
metallicity gradients. In our opinion, the large scatter in the average 
$h(X)/h(Y)$ obtained in earlier studies was due to the broad interval of 
the observed $h(X)/h(Y)$ values (Figs.~\ref{figure:f5}a--\ref{figure:f5}d).

The dependences between $h(X)/h(Y)$ and the galaxy type in the $BIK$ bands 
were studied previously in \citet{grijs98}. The plot of $h(B)/h(I)$ as a 
function of the morphological type of the galaxy in Fig.~\ref{figure:f5}d 
reproduces the results of \citet{grijs98}: $h(B)/h(I) = 1.0-1.2$ for S0 
galaxies, the minimum value $h(B)/h(I) = 1.0$ is characteristic for all 
types of galaxies, and the maximum ratios increase from 1.2-–1.4 for Sa 
galaxies to 1.6-–1.9 for Sc–-Sd galaxies.

Figures~\ref{figure:f5}a--\ref{figure:f5}c present $h(B)/h(K)$, $h(B)/h(I)$, 
and $h(I)/h(K)$ as functions of the ellipticity $e$. The ranges for the 
ratios of the linear disk scales are fairly large; the minimum values for 
$h(B)/h(K)$, $h(B)/h(I)$, and $h(I)/h(K)$ are unity independent of the 
inclination of the disk, while the maximum ratios increase with increasing 
inclination. Note that $h(I)/h(K)$ increases with the disk inclination only 
weakly. For the sample of 144 galaxies, we obtained the dependence 
$h(I)/h(K) = (1.07\pm0.03)+(0.02\pm0.10)e$. The fit parameters for the 
dependence of $h(B)/h(K)$ on $e$ have very large errors.

The correlation between $h(B)/h(I)$ and $h(I)$ noted in \citet{cunow01} is 
generally confirmed; however, the scatter of the data is very large: for 
galaxies with $h(I) \approx 1$ kpc, $h(B)/h(I) = 0.9-1.4$, while, for 
galaxies with $h(I) \approx 8$ kpc, $h(B)/h(I) = 1.0-1.7$.

\begin{figure*}
\centerline{\includegraphics[width=10cm]{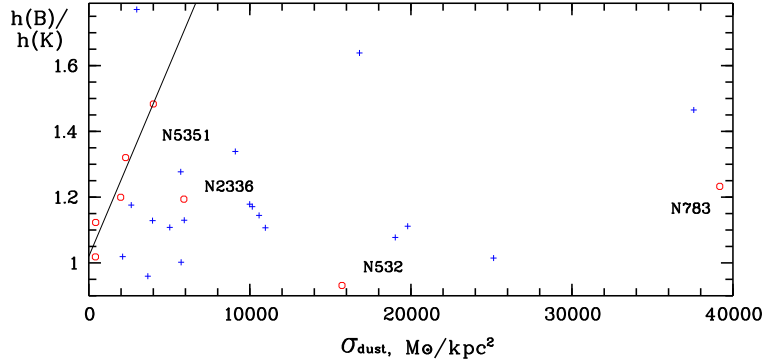}}
\caption{Ratio of the radial disk scales $h(B)/h(K)$ as functions of the 
average surface density of dust $\langle \sigma_{\rm dust} \rangle$. The 
circles denote objects from the sample of 9 galaxies, and vertical crosses 
those from the sample of 37 galaxies. An explanation for the solid line is 
given in the text. The NGC numbers of the galaxies are indicated.
\label{figure:f6}}
\end{figure*}

We also considered the dependence of $h(X)/h(Y)$ and on the average 
surface density of dust $\langle \sigma_{\rm dust} \rangle$. To this end, we 
used the dust masses derived from the FIR luminosities of 37 out of the 
144 galaxies in our sample, as well as the dust masses for 9 galaxies 
whose disk parameters were determined by us. We calculated the average 
surface density of dust using the formula 
$\langle \sigma_{\rm dust} \rangle = M_{\rm dust}/[\pi R_{25}^2 (1-e)]$. 
Figure~\ref{figure:f6} presents the resulting graph for $h(B)/h(K)$ values. 
No unambiguous dependence between the disk scale ratio and the average 
surface density of dust can be discerned. The disks of many galaxies with 
large dust densities display relatively small $h(B)/h(K)$ values 
(Fig.~\ref{figure:f6}). The data for 7 galaxies from our sample (all except 
NGC~532 and NGC~783) and for 7 galaxies with modest values from the total 
sample yield the dependence 
$h(V)/h(I) = (1.00\pm0.04)+(3.2\pm1.4)\cdot 10^{-5} 
\langle \sigma_{\rm dust}\rangle$ ($r = 0.76$), where the units for 
$\sigma_{\rm dust}$ are $M_\odot$/kpc$^2$. $h(B)/h(K)$ is more weakly 
correlated with $\sigma_{\rm dust}$: using only 5 galaxies from our sample, 
we can derive the dependence 
$h(B)/h(K) = (1.02\pm0.04)+(11.6\pm1.8)\cdot 10^{-5} 
\langle \sigma_{\rm dust} \rangle$, with $r = 0.96$ (Fig.~\ref{figure:f6}). 
Why do many galaxies with higher dust abundances display modest $h(V)/h(I)$ 
and $h(B)/h(K)$? We suggest that, when calculating the integrated dust mass 
in galaxies from observations, we cannot discriminate between dust that 
forms the exponential dust disk and dust that is concentrated towards the 
spiral arms and bars. For example, the galaxy NGC~5351, whose dust is 
concentrated in the disk, displays the highest $h(B)/h(K)$. At the same 
time, NGC~532, NGC~783, and NGC~2336, which are more dust-abundant, display 
relatively small $h(B)/h(K)$, due to the fact that a large fraction of 
their dust is concentrated in bands along the inner edges of their spiral 
arms \citep{gusev06}.

\subsection{Estimate of the central surface density of the disks}

\begin{figure*}
\centerline{\includegraphics[width=10cm]{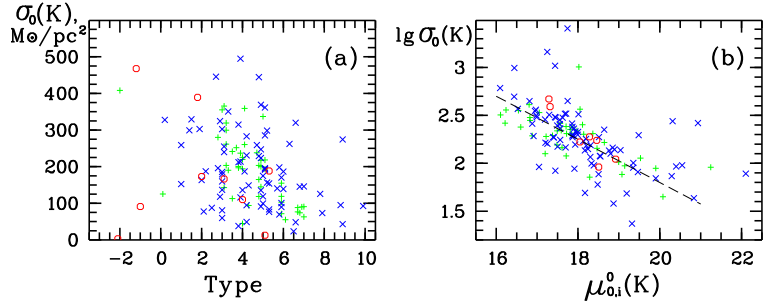}}
\caption{(a) The dependence of the central surface density of the disk 
$\sigma_0$ derived from $h(K)$ on the galaxy type. (b) Logarithm of the 
central disk surface density $\log \sigma_0(K)$ as a function of the 
central surface brightness $\mu_{0,i}^0(K)$. The dotted line in (b) 
indicates the dependence for the sample of 392 galaxies. Notation is the 
same as in Fig.~\ref{figure:f1}.
\label{figure:f7}}
\end{figure*}

Classical thin exponential disks display a dependence between their 
central surface density $\sigma_0$, linear scale length $h$, and maximum 
disk rotational velocity $V_{\rm disk}$: 
$\sigma_0 \approx 0.044V_{\rm disk}^2/h$, where $\sigma_0$ is measured in 
in $M_\odot$/pc$^2$, $V_{\rm disk}$ in km/s, and $h$ in kpc. Here, 
$V_{\rm disk} = (0.6 - 0.8)V_{\rm rot}$ (depending on the model for the galaxy), 
where $V_{\rm rot}$ is the maximum rotational velocity derived from 
observations. Thus, $\sigma_0 \approx 0.022V_{\rm rot}^2/h$. In most 
galaxies, the central disk surface density lies in the range 
50--500~$M_\odot$/pc$^2$. We can see a weak correlation between $\sigma_0(K)$ 
and the galaxy type: on average, earlier type galaxies display larger 
$\sigma_0$ values (Fig.~\ref{figure:f7}a). This is consistent with the 
dependence on galaxy type obtained for the central $K$ surface brightness 
of the disk (Fig.~\ref{figure:f1}c).

The quantities $\mu_{0,i}^0(K)$ and $\sigma_0(K)$ are fairly well correlated 
in galactic disks (Fig.~\ref{figure:f7}b). $\sigma_0(K)$ can be estimated 
from $\mu_{0,i}^0(K)$ with an accuracy of $\pm50\%$. A less clear correlation 
is observed when $\mu_{0, i}^0$ and $\sigma_0$ values measured in other 
photometric bands are considered.

\section{Conclusions}

1. In the transition from early to late type galaxies, the central $K$ 
surface brightness and the central surface density of the galactic disks 
decrease, the integrated and central color indices decrease, and the 
relative size of the disk $h/R_{25}$ and the ratio $h(X)/h(Y)$ increase 
(here, $X$ is a shorter wavelength photometric band than $Y$). The color 
gradient (normalizing by $R_{25}$) and the blue central surface brightness 
$\mu_{0,i}^0(B)$ are independent of the galaxy type. The disks in early 
type galaxies appear to be denser at the center and shorter than the disks 
in late type galaxies. The impact of dust on the photometric parameters 
of the disks and galaxies as a whole increases in the transition to late 
type galaxies.

2. The disks in S0 galaxies have more homogeneous parameters than those 
in spiral galaxies. This may be due to the lower linear age and 
metallicity gradients of their stellar populations, as well as the 
lower amounts of dust in the disks of S0 galaxies. No sharp boundary in 
the properties of disks in lenticular, spiral, and irregular galaxies 
has been found --– all parameters vary smoothly along the Hubble sequence.

3. In all photometric bands, the central surface brightnesses of the disks 
increase with the total luminosity of the parent galaxy.

4. The ratio of linear disk scales measured in different photometric bands 
$h(X)/h(Y)$ increases with the isophote ellipticity $e$ of the disk (the 
inclination of the galaxy); however, the range of $h(X)/h(Y)$ values for 
each $e$ value exceeds the range of variations of $h(X)/h(Y)$ over $e$. 
This is due to the fact that very broad intervals are observed for the 
radial variations of the composition of the stellar population in the disk 
and the parameters of the dust disks in the galaxies.

5. Assuming that the surface density distribution in the disk corresponds 
to the $K$-band surface brightness distribution, the dependence 
$h(\sigma) = h(I)/[1.07+0.02e]$ can be used to determine the linear scale 
for the decrease of the surface density $h(\sigma)$ with an accuracy of 
$\pm15\%$.

\section*{Acknowledgement}

The authors thank A.V.~Zasov (SAI) for useful discussions. This study was 
supported by the Russian Foundation for Basic Research (projects 
08--02--01323 and 10--02--91338).

\end{document}